\documentclass[12pt]{iopart_amc}

\begin{document}
\begin{flushright}
INR-TH-2016-021
\end{flushright}
\title{Instability of Static Semi-Closed Worlds in Generalized Galileon Theories}

\author[O. A. Evseev, O. I. Melichev]{O. A. Evseev, O. I. Melichev}
\address{Faculty of Physics, M~V~Lomonosov Moscow State University, Vorobyovy Gory, 1-2, Moscow, 119991, Russia}
\eads{\email{oa.evseev@physics.msu.ru}, \email{olegmelichev@gmail.com}}
\vspace{10pt}

\begin{indented}
\item[]July 2016
\end{indented}

\begin{abstract}
We consider generalized Galileon theories within general relativity in four-dimensional space-time. We provide the argument showing that the generalized Galileons described by a wide class of Lagrangians do not admit stable, static, spherically symmetric semi-closed worlds. We also show that in a class of theories with $p_{\perp} = - \rho$ (where $p_{\perp}$ is transverse pressure and $\rho$ is energy density), semi-closed worlds, if exist, would be observed as objects of negative mass.
\end{abstract}

%\pacs{???????}
%\maketitle

\section{Introduction and summary}
\label{sec:intro}

Models with Galileons are of interest, as they admit stable, null energy condition (NEC) violating solutions \cite{Nicolis:2008in, Deffayet:2009wt, Goon:2011uw, Goon:2011qf, Kobayashi:2011nu, Rubakov:2014jja}. The property of NEC-violation makes Galileons natural candidates for fields that may support Lorentzian wormholes \cite{Morris:1988cz, Morris:1988tu, wormholes_static, Hochberg:1998ha, Novikov:2007zz, Shatskiy:2008us} and/or semi-closed worlds \cite{Frolov:1988vj, Guendelman:2010pr, Chernov:2007cm, Dokuchaev:2012vc}. It has been shown, however, that asymptotically flat static and spherically symmetric wormholes are unstable in a class of generalized Galileon theories \cite{Rubakov:2015gza, Rubakov:2016zah}. The main purpose of this paper is to extend this result to semi-closed worlds.

\begin{figure}[h]
  \hfill\includegraphics[width=0.2\linewidth]{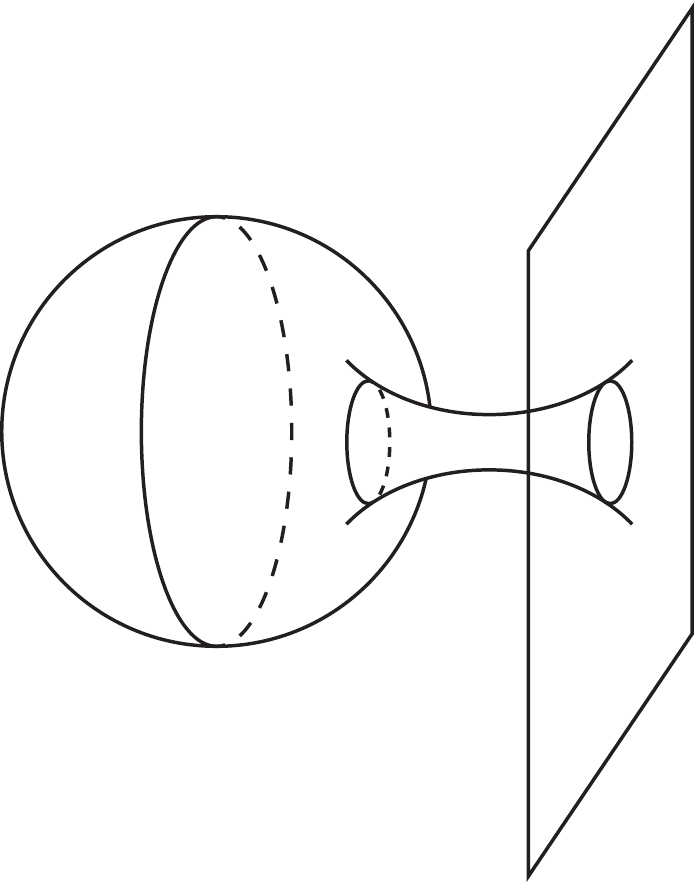}\hspace*{\fill}
  \caption{Semi-closed world.}
  \label{fig:semi-closed_world}
\end{figure}

\begin{figure}[h]
  \hfill\includegraphics[width=0.48\linewidth]{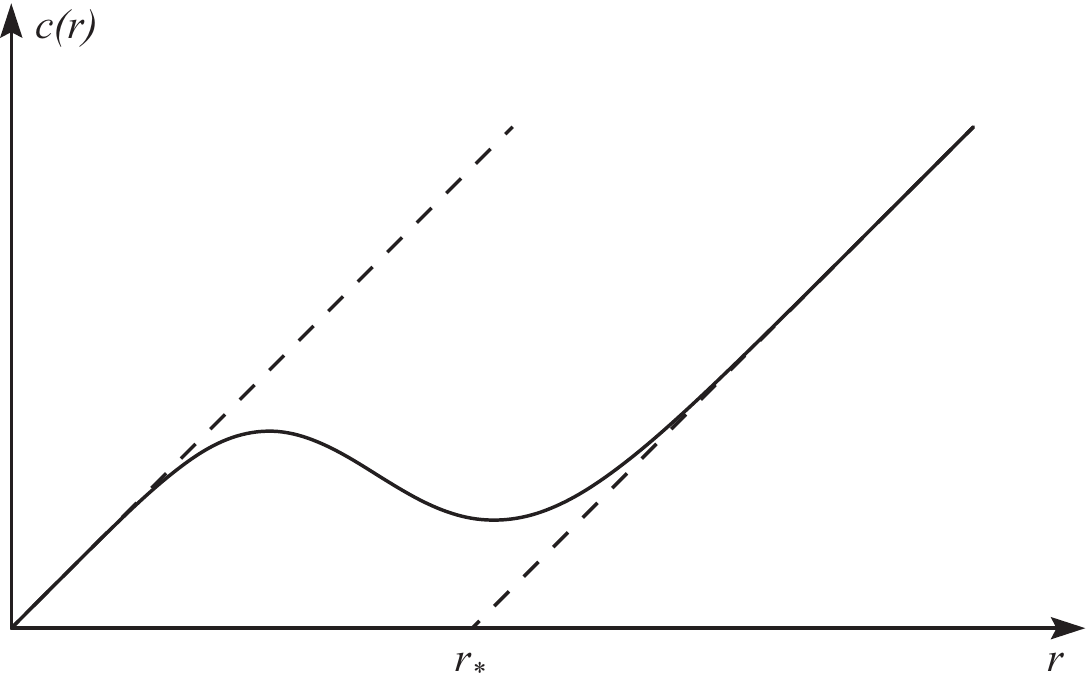}\hspace*{\fill}
  \caption{Behavior of $c(r)$ for a semi-closed world.}
  \label{fig:c}
\end{figure}

Geometry of a static, spherically symmetric, asymptotically flat semi-closed world is schematically shown in Fig.~\ref{fig:semi-closed_world}. It is described by the following metric
(signature $(+,-,-,-)$):
\begin{equation*}
\label{metric_general}
ds^2 = a^2 ( r ) dt^2 - b^2 ( r ) dr^2 - c^2 ( r ) \gamma_{\alpha \beta}dx^\alpha dx^\beta
\end{equation*}
with asymptotics
\begin{eqnarray*}
&r \rightarrow 0 \text{: }
&a \rightarrow a_0 \text{, }
b \rightarrow b_0 \text{, }
c \rightarrow b_0 r \text{, } \label{eq:asymptotics_0_general} \\
&r \rightarrow \infty \text{: }
&a \rightarrow a_\infty \text{, }
b \rightarrow b_\infty \text{, }
c \rightarrow b_\infty \left( r - r_{*} \right) \text{, } \label{eq:asymptotics_infty_general}
\end{eqnarray*}
where $\gamma_{\alpha \beta} dx^\alpha dx^\beta$ is the metric of a unit two-dimensional sphere, $a_0$, $b_0$, $a_\infty$, $b_\infty$ and $r_{*}$ are positive constants, see Fig.~\ref{fig:c}. A defining feature of a semi-closed world metric is the existence of a throat and hence non-vanishing $r_{*} > 0$.

Let us summarize our findings. Spherical symmetry dictates that the non-vanishing components of the stress-energy tensor are $T^0_0$, $T^r_r$ and $T^\alpha_\beta = T^\Omega \delta^\alpha_\beta$. In Galileon theories, as well as in many other scalar theories, the stress-energy tensor satisfies the relation $T^0_0 = T^\Omega$, \textit{i.e.} $p_{\perp} = - \rho$, where $p_{\perp}$ is transverse pressure and $\rho$ is energy density. We show that in general relativity this property alone ensures that the mass of a semi-closed world, as seen by outside observer, is \textit{negative}.

Even though negative mass objects may not be pathological in General Relativity \cite{Bondi:1957zz, Hammond:2013vqa, Belletete:2015tra}, this result calls for more detailed analysis of concrete theories. Here we specify to a generalized Galileon theory with the  Lagrangian
\begin{equation}
\label{eq:lagrangian}
\mathcal{L} = F ( \pi , X ) + K ( \pi , X ) \Box \pi \text{,}
\end{equation}
where $\pi$ is a scalar field, $F$ and $K$ are arbitrary functions and the following notation is used:
\begin{eqnarray*}
X &= \nabla_\mu \pi \nabla^\mu \pi \text{,} \\
\Box \pi &= \nabla_\mu \nabla^\mu \pi \text{.}
\end{eqnarray*}
We show that there is either a ghost or a gradient instability of perturbations about  any non-singular semi-closed world solution irrespectively of the forms of the Lagrangian functions $F$ and $K$. This is our main result: Galileons do not support static, spherically symmetric semi-closed worlds.

The paper is organized as follows. We obtain the result on the  negative mass of a semi-closed world in theories with $T^0_0 = T^\Omega$ in Sec.~\ref{sec:negative_mass}. The properties of the generalized Galileon theories with the Lagrangian~\eqref{eq:lagrangian} are discussed in Sec.~\ref{sec:galileons}. Sec.~\ref{sec:proof} gives the argument that static, spherically symmetric semi-closed worlds are unstable in the generalized Galileon theories.
\section{Negative mass}
\label{sec:negative_mass}

In what follows we use the gauge
\begin{equation}
\label{eq:gauge}
b = \frac{1}{a} \text{,}
\end{equation}
so that the semi-closed world metric is
\begin{equation*}
\label{eq:metric}
ds^2 \! = \! a^2 dt^2 - \frac{1}{a^2} dr^2 - c^2 \gamma_{\alpha \beta}dx^\alpha dx^\beta \text{,}
\end{equation*}
where $a(r)$ and $c(r)$ obey the following boundary conditions:
\begin{eqnarray*}
&r \rightarrow 0 \text{: } 
&a \rightarrow a_0 \text{, } 
c \rightarrow \frac{r}{a_0} \text{, } \label{eq:asymptotics_0} \\
&r \rightarrow \infty \text{: }
&a \rightarrow a_\infty \text{, }
c \rightarrow \frac{ r - r_{*} }{a_\infty} \text{, } \label{eq:asymptotics_infty}
\end{eqnarray*}
In this gauge the components of the Einstein tensor $G^{\mu}_{\nu} \equiv R^{\mu}_{\nu} - \frac{1}{2} \delta^{\mu}_{\nu} R$ read
\begin{eqnarray*}
G^0_0 = - 2 a^2 \Bigg[ \frac{a'c'}{ac} + \frac{1}{2} \left( \left(\frac{c'}{c}\right)^2 - \frac{1}{a^2 c^2} \right) \Bigg] - 2 a^2 \frac{c''}{c} \text{,} \label{eq:einstein_00} \\
G^r_r = - 2 a^2 \Bigg[ \frac{a'c'}{ac} + \frac{1}{2} \left( \left(\frac{c'}{c}\right)^2 - \frac{1}{a^2 c^2} \right) \Bigg]\text{, \hspace{5mm}} \label{eq:einstein_rr} \\
G^\alpha_\beta = \delta^\alpha_\beta G^\Omega \textrm{,}\phantom{\Bigg[\Bigg]} \label{eq:einstein_alpha_beta} \\
 G^\Omega = - a^2 \Bigg[ \frac{a''}{a} + \left(\frac{a'}{a}\right)^2 + \left( \frac{c''}{c} + 2 \frac{a'c'}{ac} \right) \Bigg] \text{.} \label{eq:einstein_omega}
\end{eqnarray*}
To simplify formulas below we set
\begin{equation*}
\label{eq:kappa_1}
8\pi G = 1 \text{.}
\end{equation*}

Before specifying to the generalized Galileons, let us consider any theory whose stress-energy tensor, in the spherically symmetric and static case, has the property
\begin{equation}
\label{eq:energy_T_00_eq_T_Omega}
T^0_0 = T^{\Omega} \text{,}
\end{equation}
where $T^{\Omega}$ determines the angular part, $T^\alpha_\beta = T^\Omega \delta^\alpha_\beta$. In this situation the Einstein equations imply $G^0_0 = G^{\Omega}$, \textit{i.e.}
\begin{equation*}
\label{eq:symmetry_eq}
\frac{a''}{a} + \left( \frac{a'}{a} \right)^2 - \frac{c''}{c} - \left( \frac{c'}{c} \right)^2 + \frac{1}{a^2 c^2} = 0 \text{.}
\end{equation*}
Upon the change of variables
\begin{equation*}
\gamma (r) = a^2(r) \text{, \hspace{5mm}} u (r) = c^2(r) \text{,}
\end{equation*}
the latter equation is written as
\begin{equation*}
\gamma'' u - u'' \gamma + 2 = 0 \text{.}
\end{equation*}
This equation can be used to express $\gamma(r)$ in terms of $u(r)$:
\begin{equation}
\label{eq:symmetry_solution_gamma}
\gamma(r) = u(r) \left[ \,\, \int\limits_{r}^{\infty} \frac{2 \tilde{r} - C }{u^2 ( \tilde{r} ) } d \tilde{r} + D \right] \text{,}
\end{equation}
where $C$ and $D$ are constants. We now consider the asymptotic behavior of this solution and determine $C$ and $D$.

As $r$ tends to $0$, $ u (r)$ tends to $r^2 / a_0^2$ and we find from \eqref{eq:symmetry_solution_gamma} that
\begin{equation*}
\label{eq:gamma_0}
\gamma (r) \equiv a^2(r) = a_0^2 \left[ 1 - \frac{C}{3} \frac{1}{r} + \mathcal{O}\left(r\right) \right]\text{.}
\end{equation*}
The requirement that $\gamma (r)$ is regular at $r = 0$ gives
\begin{equation*}
\label{eq:constant_C}
C = 0 \text{.}
\end{equation*}
As $r$ tends to  $\infty$, $u (r)$ tends to $(r - r_{*})^2 / a_\infty^2$ and eq.~\eqref{eq:symmetry_solution_gamma} gives
\begin{equation*}
\label{eq:gamma_infty}
\gamma (r) \equiv a^2(r) = a_\infty^2 \left[ 1 + \frac{2 r_{*} - C}{3} \frac{1}{r - r_{*}} + D (r - r_{*})^2  + o\left(\frac{1}{r - r_{*}}\right) \phantom{\int\limits_\int^\int \!\!\!\!\!\!\!\!\!\!\!} \right] \!\! \text{.}
\end{equation*}
Since $\gamma (r)$ should not grow as $r \rightarrow \infty$, we have
\begin{equation*}
\label{eq:constant_D}
D = 0 \text{.}
\end{equation*}
Thus, the relation between $\gamma(r)$ and $u(r)$ is
\begin{equation}
\label{eq:gamma_final}
\gamma(r) = u(r) \int\limits_{r}^{\infty} \frac{2 \tilde{r} }{u^2 ( \tilde{r} ) } d \tilde{r} \text{.}
\end{equation}

To find the mass of a semi-closed world as seen by an outside observer, we note that the radius of distant sphere is
\begin{equation*}
\label{eq:R_def}
R = \frac{r - r_{*}}{a_{\infty}} \text{.}
\end{equation*}
Therefore, upon the rescaling of the time coordinate
\begin{equation*}
\label{eq:T_def}
T = a_{\infty} t \text{,}
\end{equation*}
eq.~\eqref{eq:gamma_final} shows that the asymptotics of the metric as $r \rightarrow \infty$ is
\begin{equation*}
ds^2 =
\left( 1 - \frac{R_g}{R} \right) dT^2
 - \frac{ 1 } { 1 - \frac{R_g}{R}} dR^2
 - R^2 \gamma_{\alpha \beta}dx^\alpha dx^\beta \text{,}
\end{equation*}
where
\begin{equation}
\label{eq:grav_radius}
R_g = - \frac{2}{3} \frac{r_{*}}{a_{\infty}}
\end{equation}
Since $r_{*} > 0$, see Fig.~\ref{fig:c}, we conclude that in theories obeying \eqref{eq:energy_T_00_eq_T_Omega}, semi-closed worlds, if exist, would be observed as objects of negative mass.

\section{Generalized Galileon theory}
\label{sec:galileons}

\subsection{Stress-energy tensor}
\label{subsec:stress_energy}

The stress-energy tensor for the theory \eqref{eq:lagrangian} is
\begin{eqnarray*}
T_{\mu \nu} =&
2 F_X \cdot \partial_\mu \pi \partial_\nu \pi
+ 2 K_X \Box \pi \cdot  \partial_\mu \pi \partial_\nu \pi \nonumber \\
&- \partial_\mu K \partial_\nu \pi
- \partial_\nu K \partial_\mu \pi
- g_{\mu \nu} F
+ g_{\mu \nu} g^{\lambda \rho} \partial_\lambda K \partial_\rho \pi \text{,}
\end{eqnarray*}
which in the static spherically symmetric case and in the gauge \eqref{eq:gauge} becomes

\begin{eqnarray*}
T^0_0 &= - F - \left( a \pi' \right)^2 K_\pi + 2 a \left( a \pi' \right)^2 \left( a \pi' \right)' K_X \text{,} \label{eq:energy_00}\\
T^r_r &= - F + \left( a \pi' \right)^2 K_\pi + 2 \left( a \pi' \right)^3 \left( \frac{a'}{a} + 2 \frac{c'}{c} \right) K_X  - 2 \left( a \pi' \right)^2 F_X \text{, } \label{eq:energy_rr} \\
 T^\alpha_\beta &= \delta^\alpha_\beta T^\Omega \text{,}  \label{eq:energy_alpha_beta} \\
 T^\Omega &= - F - \left( a \pi' \right)^2 K_\pi + 2 a \left( a \pi' \right)^2 \left( a \pi' \right)' K_X \text{.} \label{eq:energy_omega}
\end{eqnarray*}
 Note that it has the property \eqref{eq:energy_T_00_eq_T_Omega}, so the result of Sec.~\ref{sec:negative_mass} applies.

\subsection{Stability conditions}
\label{subsec:stability}

The perturbations about static, spherically symmetric background ($\pi \rightarrow \pi + \chi$) are described by the following effective quadratic Lagrangian \cite{Rubakov:2015gza} (in the gauge \eqref{eq:gauge})
\begin{equation*}
\mathcal{L}^{(2)} = a^{-2} \mathcal{G}^{00} \dot{\chi}^2 - a^{2} \mathcal{G}^{rr} \left( \chi ' \right)^2 - c^{-2}  \mathcal{G}^{\Omega} \gamma^{\alpha \beta} \partial_\alpha \chi \partial_\beta \chi \text{,}
\end{equation*}
where the effective metric is
\begin{eqnarray*}
\mathcal{G}^{00}     &= F_X - K_\pi - a^2 K'_X \pi' - 2 a K_X \left( a \pi' \right)' - 4 a^2 K_X \frac{c'}{c} \pi' - K_X^2 \left( a \pi' \right)^4 \\
\mathcal{G}^{\Omega} &= F_X - K_\pi - a^2 K'_X \pi' - 2 a K_X \left( a \pi' \right)' \nonumber \\
&- 2 a^2 K_X \frac{c'}{c} \pi' - 2 a a' K_X \pi' - K_X^2 \left( a \pi' \right)^4 \\
\mathcal{G}^{rr} &= F_X - 2 F_{XX} \left( a \pi' \right)^2 - K_\pi + a^2 K'_X \pi' - 2 a^2 K_X \pi' \left( \frac{a'}{a} + 2 \frac{c'}{c} \right) \nonumber \\
&+ 2 a K_{XX} \left( a \pi' \right)^2  \left( a \pi' \right)' + 2 a K_{XX} \left( a \pi' \right)^3  \left( \frac{a'}{a} + 2 \frac{c'}{c} \right) + 3 K_X^2 \left( a \pi' \right)^4
\end{eqnarray*}
and terms without derivatives of $\chi$ are omitted. Hereafter $\pi$ denotes the background field. The background is stable as long as
\begin{equation}
\label{eq:stability}
\mathcal{G}^{00} > 0 \text{, }
\mathcal{G}^{\Omega} \ge 0 \text{, }
\mathcal{G}^{rr} \ge 0 \text{. }
\end{equation}
From now on we are interested in the first two of these conditions. Using the combination of the Einstein equations $G^0_0 - G^r_r = T^0_0 - T^r_r$, that is
\begin{equation*}
\label{eq:G_00_rr_eq_T_00_rr}
- 2 a^2 \frac{c''}{c} = 2 \left( a \pi' \right)^2 \left[ F_X - K_\pi + a K_X \left( a \pi' \right)' - a^2 K_X \pi' \left( \frac{a'}{a} + 2 \frac{c'}{c} \right) \right] \text{,}
\end{equation*}
we write
\begin{eqnarray}
\left(\pi' \right)^2 \mathcal{G}^{00} &= - \left ( \frac{c'}{c} + a^2 K_X  \left(\pi' \right)^3 \right)' - \left ( \frac{c'}{c} + a^2 K_X  \left(\pi' \right)^3 \right)^2 \text{,} \label{eq:G_00_grouped}\\
\left(\pi' \right)^2 \mathcal{G}^{\Omega} &= \mathcal{G}^{00}  \left(\pi' \right)^2 + 2  a^2 K_X  \left(\pi' \right)^3 \left( \frac{c'}{c} - \frac{a'}{a} \right)  \label{eq:G_Omega_grouped} \text{,}
\end{eqnarray}
It is now natural, following \cite{Rubakov:2016zah}, to introduce the variable
\begin{equation*}
Q = \frac{c'}{c} + a^2 K_X  \left(\pi' \right)^3
\end{equation*}
and rewrite eqs.~\eqref{eq:G_00_grouped}, ~\eqref{eq:G_Omega_grouped} as follows:
\begin{eqnarray*}
\left(\pi' \right)^2 \mathcal{G}^{00} &= -Q' - Q^2, \label{eq:g_00_with_Q} \\
\left(\pi' \right)^2 \mathcal{G}^{\Omega} &= -Q' - Q^2 + 2 \left( Q - \frac{c'}{c} \right) \left( \frac{c'}{c} - \frac{a'}{a} \right) \text{.} \label{eq:g_Omega_with_Q}
\end{eqnarray*}

Since $c'/c$ tends to $1/r$ both as $r \rightarrow 0$ and as $r \rightarrow \infty$, it is convenient to introduce the variable
\begin{equation*}
q = Q - \frac{1}{r} \text{.}
\end{equation*}
The expressions for $\mathcal{G}^{00}$ and $\mathcal{G}^{\Omega}$ in terms of $q$ are
\begin{eqnarray*}
\left(\pi' \right)^2 \mathcal{G}^{00} &= -q' - \frac{2q}{r} - q^2, \label{eq:g_00_with_q} \\
\left(\pi' \right)^2 \mathcal{G}^{\Omega} &= -q' - \frac{2q}{r} - q^2 + 2 \left( q + \frac{1}{r} - \frac{c'}{c} \right) \left( \frac{c'}{c} - \frac{a'}{a} \right) \text{,} \label{eq:g_Omega_with_q}
\end{eqnarray*}

\section{Proof of instability}
\label{sec:proof}

The main idea of the proof below is to show that $q$ is negative at $r = 0$ and positive at $r = \infty$ and, using these properties, show that the stability conditions \eqref{eq:stability} are violated for any non-singular configuration.

\subsection{The sign of \texorpdfstring{$q$}{q} at \texorpdfstring{$r = 0$}{r = 0}}
\label{subsec:q_at_0}
We are going to prove that $q$ is negative at $r = 0$. To this end, let us assume the opposite,
\begin{equation}
\label{eq:q_g_0_at_0}
\left. q \right|_{r \rightarrow 0} > 0 \text{.}
\end{equation}
We now distinguish three types of the asymptotic behavior of $q$ at $r = 0$. First, let us consider the case when $q$ grows, as $r \rightarrow 0$, faster than $1/r$:
\begin{equation}
\label{eq:q_gg_1_over_r_at_0}
q \gg \frac{1}{r} \text{.}
\end{equation}
This leads to
\begin{equation}
\label{eq:G_00_at_r_0}
\left(\pi' \right)^2 \mathcal{G}^{00} \approx -q' - q^2 > 0 \text{.}
\end{equation}
We now integrate the inequality in \eqref{eq:G_00_at_r_0} from $0$ to $r$ and obtain
\begin{equation*}
\frac{1}{q} > r \text{,}
\end{equation*}
which contradicts \eqref{eq:q_gg_1_over_r_at_0} under the assumption~\eqref{eq:q_g_0_at_0}. This means that $q$, if positive, cannot grow faster than $1/r$.

The second case to consider is $q \sim 1/r$:
\begin{equation*}
\label{eq:q_sim_1_over_r_at_0}
\left. q \right|_{r \rightarrow 0} = \frac{\alpha}{r} + o \left( \frac{1}{r} \right) \text{,}
\end{equation*}
where $\alpha > 0$ because of \eqref{eq:q_g_0_at_0}. In this case we have
\begin{equation}
\left(\pi' \right)^2 \mathcal{G}^{00} = \frac{ - \alpha ( 1 + \alpha ) }{r^2} + o \left( \frac{1}{r^2} \right) \text{.} \label{eq:g_00_when_q_sim_1_over_r}
\end{equation}
The right hand side of~\eqref{eq:g_00_when_q_sim_1_over_r} is negative, meaning that positive $q \sim 1/r$ is also impossible.

Let us finally consider the case
\begin{equation}
\label{eq:q_ll_1_over_r_at_0}
\left. q \right|_{r \rightarrow 0} = o \! \left( r^{-1} \right) \text{,}
\end{equation}
which leads to
\begin{equation}
\label{eq:g_00_with_small_q}
\left(\pi' \right)^2 \mathcal{G}^{00} \approx -q' - \frac{2q}{r} > 0 \text{.}
\end{equation}
Under the assumption of positive $q$ this is the equivalent to
\begin{equation}
\label{eq:dq_over_q}
-\frac{dq}{q} > 2\frac{dr}{r} > \frac{dr}{r} \text{.}
\end{equation}
We may now choose $r_1$ and $r_2 > r_1$ so that $q > 0$ in the whole interval $[r_1, r_2]$ and integrate the inequality~\eqref{eq:dq_over_q} from $r_1$ to $r_2$. We get
\begin{equation}
\label{eq:integral_result}
q_1 r_1 > q_2 r_2 \text{.}
\end{equation}
The condition~\eqref{eq:g_00_with_small_q} under the assumption~\eqref{eq:q_g_0_at_0} requires $q'$ to be negative, meaning that $q$ is monotonous by decreasing, which leads to the contradiction between \eqref{eq:integral_result} and \eqref{eq:q_ll_1_over_r_at_0}.

Thus, the arguments above prove that $q < 0$ near $r = 0$.

\subsection{The sign of \texorpdfstring{$q$}{q} at \texorpdfstring{$r = \infty$}{r = infinity}}
\label{subsec:q_at_infty}
We needed only $G^{00}$ to prove that $q$ is negative near $r = 0$. To prove its positivity at $r = \infty$ we also need $G^{\Omega}$. Since $c \rightarrow r - r_{*}$ as $r \rightarrow \infty$, we have
\begin{equation}
\frac{c'}{c} = \frac{1}{r} + \frac{r_{*}}{r^2} + o\left(\frac{1}{r^2}\right) \text{.}\label{eq:assumption_c_infty}
\end{equation}
Now, we have $a \rightarrow a_\infty \left[ 1 - M/ \left(r - r_{*}\right) \right]$ as $r \rightarrow \infty$ (with negative $M$, see \eqref{eq:grav_radius}), hence
\begin{equation}
\frac{a'}{a} = \mathcal{O} \left(\frac{1}{r^2}\right) \text{.} \label{eq:assumption_a_infty}
\end{equation}
Using \eqref{eq:assumption_c_infty} and \eqref{eq:assumption_a_infty} we obtain
\begin{equation*}
\left(\pi' \right)^2 \mathcal{G}^{\Omega} = -q' - q^2 - \frac{2r_{*}}{r^3} + \mathcal{O} \left(\frac{q}{r^2}\right) + o\left(\frac{1}{r^3}\right) \\
\end{equation*}
We now prove that $\left. q \right|_{r \rightarrow \infty} > 0$. Let us assume the opposite:
\begin{equation}
\label{eq:q_l_0_at_infty}
\left. q \right|_{r \rightarrow \infty} < 0 \text{.}
\end{equation}
Let us again consider the three types of asymptotic behavior of $q$.

The first case is that as $r \rightarrow \infty$
\begin{equation*}
\label{eq:q_gg_1_over_r_at_infty}
\left| q \right| \gg \frac{1}{r} \text{,}
\end{equation*}
which leads to
\begin{equation*}
\label{eq:G_00_at_r_infty}
\left(\pi' \right)^2 \mathcal{G}^{00} \approx -q' - q^2 > 0 \text{.}
\end{equation*}
Integrating $q'/q^2 < -1$ from $r_1$ to $r_2 > r_1$ we obtain
\begin{equation*}
\frac{1}{q(r_2)} > \frac{1}{q(r_1)} + r_2 - r_1 \text{,}
\end{equation*}
Since $q(r_1) < 0$ by assumption, and $r_2$ can be arbitrarily large, there is a singularity point at which $1/q(r_2) = 0$. This shows that $q < 0$ cannot tend to zero slower than $1/r$.

The second case to consider is
\begin{equation*}
\label{eq:q_sim_1_over_r_at_infty}
\left. q \right|_{r \rightarrow \infty} = \frac{\alpha}{r} + o \left( \frac{1}{r} \right) \text{,}
\end{equation*}
where $\alpha < 0$  under the assumption \eqref{eq:q_l_0_at_infty}. This leads to
\begin{equation}
\left(\pi' \right)^2 \mathcal{G}^{\Omega} = \frac{ \alpha ( 1 - \alpha ) }{r^2} + o \left( \frac{1}{r^2} \right) \text{.} \label{eq:g_Omega_when_q_sim_1_over_r_at_infty}
\end{equation}
The right hand side of~\eqref{eq:g_Omega_when_q_sim_1_over_r_at_infty} is negative, so the background is unstable.

Finally, we consider the case
\begin{equation}
\label{eq:q_ll_1_over_r_at_infty}
\left. q \right|_{r \rightarrow \infty} = o \! \left( r^{-1} \right) \text{.}
\end{equation}
Then the expression for $\mathcal{G}^{\Omega}$ is
\begin{equation}
\label{eq:G_Omega_at_infty}
\left(\pi' \right)^2 \mathcal{G}^{\Omega} = -q' - q^2 - \frac{2r_{*}}{r^3} + o\left(\frac{1}{r^3}\right) \text{.}
\end{equation}
Under the assumptions \eqref{eq:q_l_0_at_infty} and \eqref{eq:q_ll_1_over_r_at_infty} there is a region where $q' > 0$: $q$ tends to $0$ from below, so that it grows at least somewhere. In this region the dominant part of $\mathcal{G}^{\Omega}$ is given by three negative terms in eq.~\eqref{eq:G_Omega_at_infty}, which contradicts the stability.

The arguments above show that $q$ has to be positive as $r$ tends to $\infty$. We have also shown that it has to be negative at $r = 0$. This means that there is a point $r_n$ where $q(r_n) = 0$, $q' (r_n) \geq 0$. The right hand side of eq.~\eqref{eq:g_00_with_q} is non-positive at $r_n$ which contradicts the stability conditions \eqref{eq:stability}. This brings us to the conclusion that the stable, static, spherically symmetric semi-closed worlds do not exist in four-dimensional Galileon theories with the Lagrangians of the form \eqref{eq:lagrangian}.

\ack

The authors are indebted to S.~Mironov and V.~Rubakov for helpful discussions. This work has been supported by Russian Science Foundation, grant 14-22-00161.

\end{document}